# Collective dark states controlled transmission in plasmonic slot waveguide with a stub coupled to a cavity dimer

Zhenzhen Liu • Jun-Jun Xiao • Qiang Zhang • Xiaoming Zhang • Keyu Tao

**Abstract**: We report collective dark states controlled transmission in metal-dielectric-metal waveguides with a stub coupled to two twin cavities, namely, plasmonic waveguide-stub-dimer systems. In absence of one individual cavity in the dimer, plasmon induced transparency (PIT) is possible when the cavity and the stub have the same resonance frequency. However, it is shown that the hybridized modes in the dimer collectively generate two dark states which make the stub-dimer "invisible" to the straight waveguide, splitting the original PIT peak into two in the transmission spectrum. Simultaneously, the original PIT peak becomes a dip due to dark state interaction, yielding anti-PIT-like modulation of the transmission. With full-wave electromagnetic simulation, we demonstrate that this transition is controlled by the dimer-stub separation and the dimer-stub relative position. All results are analytically described by the temporal coupled mode theory. Our results may be useful in designing densely integrated optical circuits, and in optical sensing and switching applications.

Z. Liu • J. J.   Xiao* • Q. Zhang • X. M. Zhang

College of Electronic and Information Engineering, Shenzhen Graduate School, Harbin Institute of Technology, Shenzhen 518055, China

Email: eiexiao@hitsz.edu.cn

K. Y. Tao

Shenzhen Key Laboratory of Micro-Nano Photonic Information Technology, THz Technical Research Center of Shenzhen University, College of Electronic Science and Technology, Shenzhen 518067, China and Department of Mechanical Engineering, University of Colorado Boulder, Boulder, CO 80309, USA

**Introduction**

Using optical bright and dark states and their mutual interactions to manipulate light in nanoscale has enabled many applications in nanophotonics. Fano resonance and electromagnetically induced transparency (EIT) are two examples.[1] Fundamentally, both of them are coherent interference processes found in quantum systems but have been observed in various classical physics.[2,3] Many optical and photonic artificial structures have been proposed to mimic EIT.[4] In particular, tremendous Fano resonance and EIT analogies were demonstrated in plasmonic clusters and metamaterials.[5-9]

Plasmonic structures offer significant manipulations of light in the nanometric scale, enabling deep-subwavelength confinement. To achieve highly flexible spectral and spatial control over light, a lot of designs have been proposed. One category of schemes relies on the usage of single or multiple resonators that are directly or evanescently coupled to a waveguide.[10] Plasmonic waveguides can be readily fabricated with metal-dielectric-metal (MDM) slots. The resonator designs, however, endure much more degrees of freedom, in light of plasmonic mode hybridization. Depending on the geometry configuration and collective excitation of resonance modes in coupled resonators, single- and multi-band filters with add/drop functionalities have been reported.[11-14] These properties are largely governed by the coherent dynamics of plasmonic wave interferences, featuring Fano line shape or plasmon induced transparency (PIT). In this vein, plasmonic waveguide-resonator system not only provides feasible way for dense on-chip photonic interconnection, but also represents a good platform to explore analogous quantum effects in multi-level atomic physics.

In this Letter, we theoretically and numerically study a plasmonic waveguide system consisting of a stub that connects to a slot waveguide and two cascaded cavity resonators (see Fig. 1). It is shown that such waveguides exhibit spectral responses similar to a plasmonic nanorod clusters that mimic atomic four-level configuration.[15] We analysis the coupling dynamics by the temporal coupled mode theory (TCMT),[16,17] and show that the proposed structure leads to a cancellation of the PIT peak, which is referred to as anti-PIT. This is similar to transforming a PIT to plasmon induced absorption (PIA).[8,9] A specific MDM structure is designed to examine this transformation in the near-infrared. Full wave simulations are then used to characterize the transmission and the near-field interactions in the proposed anti-PIT structure.

**Theoretical Analysis**

Let us start from the general scheme of the proposed anti-PIT structure, as shown in Fig. 1. The system basically consists of a "bright" resonator $|0\rangle$ interacting with a continuum $|c\rangle$ and two discrete states $|1\rangle$ and $|2\rangle$. The twin discrete states are supported by two closely coupled cavities, bearing bonding and anti-bonding hybridized modes. Within the TCMT, the amplitudes of the incoming and outgoing waves in the waveguide are denoted by $S_+$ and $S_-$, respectively. Here we consider light coming from one port and $S_\pm$ are normalized such that $|S_\pm|^2$ represent the power of the wave. The time-harmonic amplitudes $a_i$ $(i=0,1,2)$ of the stub resonator and the two cavities are expressed as:

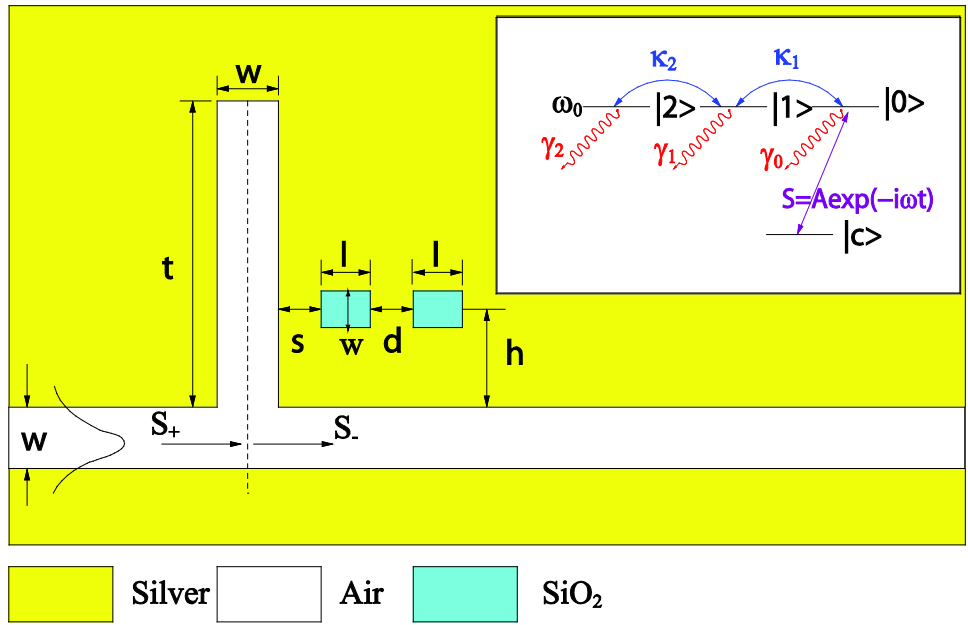

FIG. 1. Schematic figure of the proposed plasmonic waveguide-resonator system. Inset shows the energy level diagram with near-field coupling.

$$\frac{da_0}{dt} = (j\omega_0 - \gamma_0 - \gamma_e)a_0 + e^{j\theta}\sqrt{\gamma_e}S_+ - j\kappa_1 a_1, \tag{1}$$

$$\frac{da_1}{dt} = (j\omega_1 - \gamma_1)a_1 - j\kappa_1 a_0 - j\kappa_2 a_2, \tag{2}$$

$$\frac{da_2}{dt} = (j\omega_2 - \gamma_2)a_2 - j\kappa_2 a_1, \tag{3}$$

where $\omega_i$ $(i=0,1,2)$ represents the resonant frequency of the $i$-th resonator and $\gamma_i$ (i=0,1,2) is the decay rate due to the internal loss. In Eqs. (1)-(3), $\gamma_e$ is the decay rate due to the energy escape into the slot waveguide, $\theta$ represents the phase of waveguide-stub coupling, $\kappa_1$ denotes the coupling strength between the stub and the dimer, and $\kappa_2$ is the coupling strength between the two cavities in the dimer. The corresponding atomic level diagram is shown in the inset of Fig. 1. By power conservation and the time-reversal symmetry, the amplitude in the stub and the outgoing and incoming waves in the waveguide must be related as

$$S_- = S_+ - e^{-j\theta}\sqrt{\gamma_e}a_0, \tag{4}$$

The waveguide transmission can be derived from Eqs. (1)-(4) as

$$T = \left|\frac{S_-}{S_+}\right|^2 = \left|1 - \frac{\gamma_e}{j(\omega-\omega_0)+\gamma_0+\gamma_e+A}\right|^2, \tag{5}$$

where $A = \kappa_1^2/[j(\omega-\omega_1)+\gamma_1+B]$, and $B = \kappa_2^2/[j(\omega-\omega_2)+\gamma_2]$. Notice that as the waveguide-stub and stub-dimer couplings are in geometrically cascading form, the transmission expression also appears in a cascading form. Equation (5) degenerates to the tooth-shaped waveguide[13] when $\kappa_1 = 0$ and becomes the case (PIT-structure) studied by Zhang et al[10] if $\kappa_2 = 0$. The PIT-structures have transmission dips at $\omega = (\omega_0 + \omega_1 \pm \sqrt{(\omega_0-\omega_1)^2 + 4\kappa_1^2})/2$. In presence of a second dark state (i.e., $\kappa_2 \neq 0$ and $B \neq 0$), the PIT-like transparency peak ($\omega = \omega_1$) splits into two at $\omega = (\omega_1 + \omega_2 \pm \sqrt{(\omega_1-\omega_2)^2 + 4\kappa_2^2})/2$ in the limit of $\gamma_{0,1,2} \to 0$, yielding double PITs.

To simplify the theoretical analysis, we first neglect the scattering radiation and the Ohmic loss in the system, i.e., let $\gamma_i = 0$ ($i = 0,1,2$), and assume that the resonance frequencies of the individual resonators are equal, $\omega_0 = \omega_1 = \omega_2$. In such case, Eq. (5) reduces to

$$T = 1 - \frac{\gamma_e^2 (\kappa_2 + \omega - \omega_0)^2 (\kappa_2 - \omega + \omega_0)^2}{\gamma_e^2 (\kappa_2 + \omega - \omega_0)^2 (\kappa_2 - \omega + \omega_0)^2 + (\omega - \omega_0)^2 (\kappa_1^2 + \kappa_2^2 - \omega^2 + 2\omega\omega_0 - \omega_0^2)^2}. \quad (6)$$

It is easy to show that the transmission dips are now at $\omega = \omega_0$ and $\omega = \omega_0 \pm \sqrt{\kappa_1^2 + \kappa_2^2}$, and the peak positions at $\omega = \omega_0 \pm \kappa_2$. Obviously there are two peaks (unity transmission) and three dips (zero transmission), which is controlled by the number of collective dark states that are theoretically predictable.[18] In this regard, the second term in the right hand side of Eq. (6), $1-T$, is equivalent to extinction in plasmonic clusters.[8,9] We note that dark states here are modes/excitations that do not affect the straight waveguide optically, the signature of which is unity transmission. This represents the situation that the side cavities collectively appear "invisible" to the waveguide mode in the slot.

Figure 2 shows the transmission spectra [by Eq. (6)] as a function of the coupling strength ($\kappa_1$, $\kappa_2$) and detuning $\Delta = \omega - \omega_0$. Figure 2(a) is for fixed $\kappa_2 = 0.096\omega_0$ and Fig. 2(b) is for fixed $\kappa_1 = 0.048\omega_0$. It is seen that the system supports three transmission dips except for vanishing $\kappa_1$ and/or $\kappa_2$. Apparently, vanishing $\kappa_2$ features a PIT-PIA-like transition. Moreover, with increasing $\kappa_1$ in Fig. 2(a) or with increasing $\kappa_2$ in Fig. 2(b), the two side dips split with one red shifted and the other blue shifted. The middle stopband slightly shrinks in Fig. 2(a), while it widens in Fig. 2(b) for increased coupling strength. The bandwidth $\delta\omega \sim \sqrt{\kappa_1^2 + \kappa_2^2} - \kappa_2$ of the two transparency windows becomes wider (narrower), with increasing coupling strength $\kappa_1$ ($\kappa_2$). For clarity, several cut lines in Figs. 2(a) and 2(b) are shown in Figs. 2(c) and 2(d), respectively.

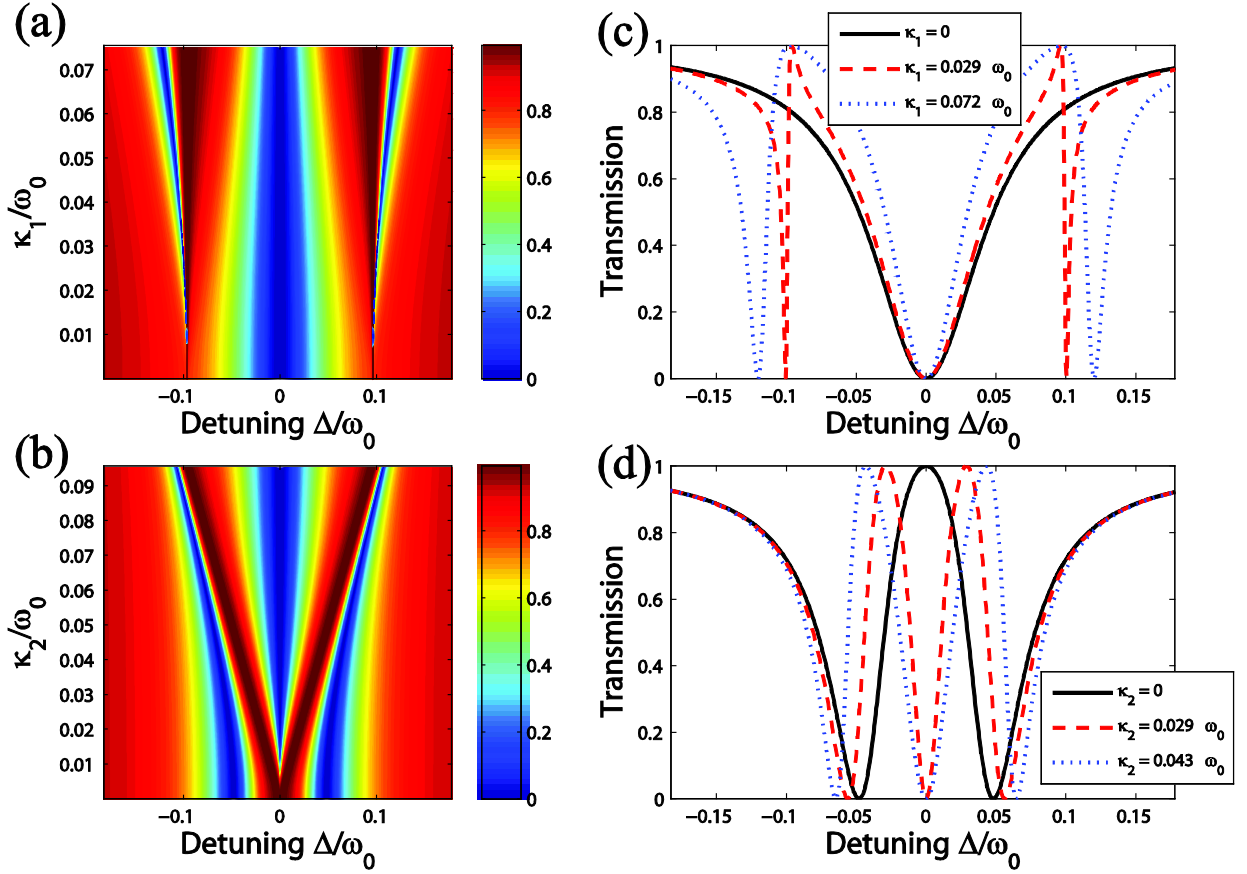

FIG. 2. Theoretical transmission dependence on $\kappa_1$, $\kappa_2$, and $\omega$ for $\gamma_e = 0.048\omega_0$. (a) $\kappa_2 = 0.096\omega_0$ and (b) $\kappa_1 = 0.048\omega_0$. (c) and (d) show the transmission spectra for $\kappa_1 = 0$, $0.029\omega_0$ and $0.072\omega_0$ in (a), and for $\kappa_2 = 0$, $0.029\omega_0$, and $0.043\omega_0$ in (b), respectively.

**Numerical Results and Discussion**

Next, we numerically study an explicit case of Fig. 1 to verify the theoretical predictions. The stub (tooth-shaped resonator) of length $t = 782$ nm is connected to the slot waveguide at one side. Two closely arranged cavities of equal length $l = 100$ nm are evanescently side-coupled to the stub. All of them are of thickness $w = 50$ nm. The distance $h = 175$ nm is used to avoid direct coupling between the dimer and the slot waveguide. The surface-to-surface distances between the stub and the two cavities are $s = 10$ nm and $d = 10$ nm, respectively. The dielectric medium in the slot and the stub waveguides is air while the two cavities are filled by SiO$_2$ ($\varepsilon = 4.5$). The metallic cladding is silver with permittivity described by the Drude model, $\varepsilon_m(\omega) = \varepsilon_\infty - \omega_p / \omega(\omega + j\Gamma)$, where $\varepsilon_\infty = 3.7$, $\omega_p = 9.1$ eV, and $\Gamma = 0.018$ eV.[11,13] Finite element method (Comsol Multiphysics) is employed to calculate the transmission with TM polarization ($E_z \equiv 0$).

Figure 3(a) shows the transmission spectrum of the proposed anti-PIT structure for both lossy and lossless (i.e., setting $\Gamma = 0$) cases. Superimposed in Fig. 3(a) is the result (dotted red) for the case with the second cavity removed (i.e., the corresponding PIT structure).[10] Presence of loss in the metallic medium simply makes the total transmission and total reflection imperfect (i.e., not strictly approaching 0 and 1) without much shifting of the peaks and the dips. We note that with only one cavity coupled to the stub, the structure exhibits EIT-like phenomena[10,11] and the transmission maximizes near the wavelength $\lambda = 900$ nm. Further notice that both the waveguide-stub and waveguide-cavity systems have minimum transmission for $\lambda \approx 900$ nm, shown by the solid black and dotted red lines in Fig. 3(b), respectively. Indeed, our structure is actually formed by introducing an additional cavity adjacent to the first one in a PIT-structure. The two cavities as a whole is a dimer and the transmission peak at $\lambda = 900$ nm falls down to zero, introducing two transparency windows [solid black in Fig. 3(a)]. This resembles the transition from PIT into PIA in judiciously arranged plasmonic clusters.[8] Due to the interference of the modes in the stub and the dimer, the bandwidth becomes narrower with increased Q factor. Figures 4(a)-4(c) show the field distribution of $H_z$ for transmission dip wavelength [marked by arrows in Fig. 3(a)] of $\lambda = 780$ nm, $898$ nm, and $1048$ nm, respectively. The transmission with or without loss are obviously different for the two side bands but are basically the same for the anti-PIT band. Figures 4(d) and 4(e) show the field pattern for the transmission dips of the PIT-structure, highlighting in-phase and out-of-phase coupling between the stub and the cavity. Figure 3(b) shows the transmission spectra for the cases with the resonant structures (stub, one cavity, and the dimer) directly coupled to the slot waveguide. The corresponding field patterns shown in Figs. 4(f) and 4(g) demonstrate that the dimer supports anti-bonding and bonding resonant modes at $\lambda = 780$ nm and $\lambda = 1030$ nm. Figures 4(h) and 4(i) show that a transmission dip at $\lambda = 900$ nm can be produced by the stub (the second-order resonance) or by the single cavity (the dipolar resonance). We note that the tooth-shaped structure in Fig. 4(h) may generate additional transmission dips by its first- and third-order modes[13] at $\lambda = 1480$ nm and 656 nm, respectively.

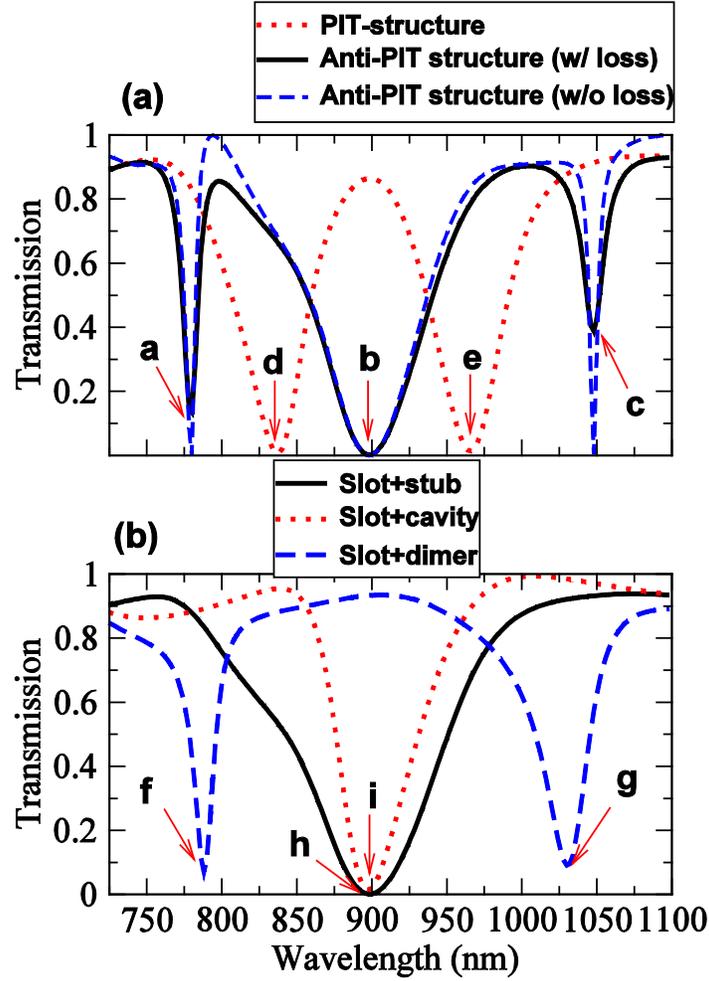

FIG. 3. (a) Transmission spectrum calculated by FEM for lossy (solid black) and lossless (dashed blue) cases of the specific waveguide-stub–dimer system (Anti-PIT structure). Transmission for the corresponding PIT structure (dotted red) is also shown. (b) Transmission spectrum of the slot waveguide side coupled to a cavity ("Slot+cavity", dashed red), a stub ("Slot+stub", solid black) and a cavity dimer ("Slot+dimer", dashed blue).

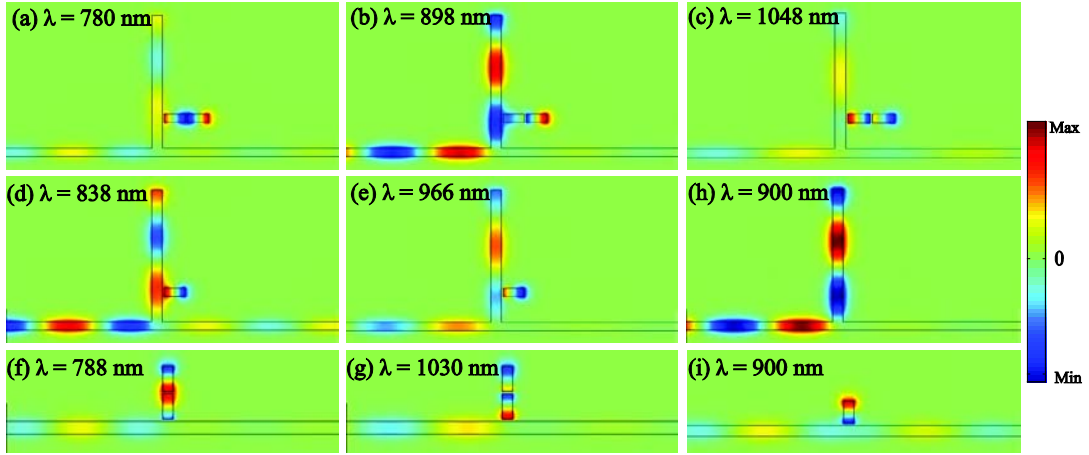

FIG. 4. Field distributions $H_z$ at the specific wavelengths marked in Figs. 3(a) and 3(b).

We are now in a position to discuss the influence of the gap $d$ and the distance $h$. Firstly, we examine the coupling strength in the dimer as shown in Fig. 5(a). The FEM simulation results are fitted by the TCMT formula Eq. (5) and they are in good agreements. Figure 5(c) shows that the extracted coupling constant $\kappa_2$ exponentially decreases as the gap $d$ between the two cavities increases from 10 nm to 80 nm, $\kappa_2 = \alpha \exp(-d/\beta)$, here $\alpha = 813.6$ THz and $\beta = 437.8$ nm. This is reasonable since the coupling inside the dimer is dominated by near-field interactions. However, in Fig. 5(a), the dips shift asymmetrically for increasing $d$, meaning that the coupling strength $\kappa_1$ and the decay rates $\gamma_i$ have also been changed even though we just vary $d$ and keep the other geometrical parameters intact. This is not unreasonable since the dimer cavities can be regarded as a whole that couples to the stub. The interactions of the elements with ($\kappa_1$, $\kappa_2$, $\gamma_i$) eventually results in PIT phenomenon in absence of the second cavity, for $\kappa_2 \approx 0$ at $d \geq 80$ nm.

Figure 5(b) shows the influence of $h$ which measures the position of the dimer to the slot waveguide. It is seen that the depth of the left and right dips vary in unequal phases when $h$ increases. Here the stub-dimer distance and the dimer gap are maintained at $s = 10$ nm and $d = 10$ nm, respectively. The coupling strength $\kappa_1$ between the stub and the dimer is governed not only

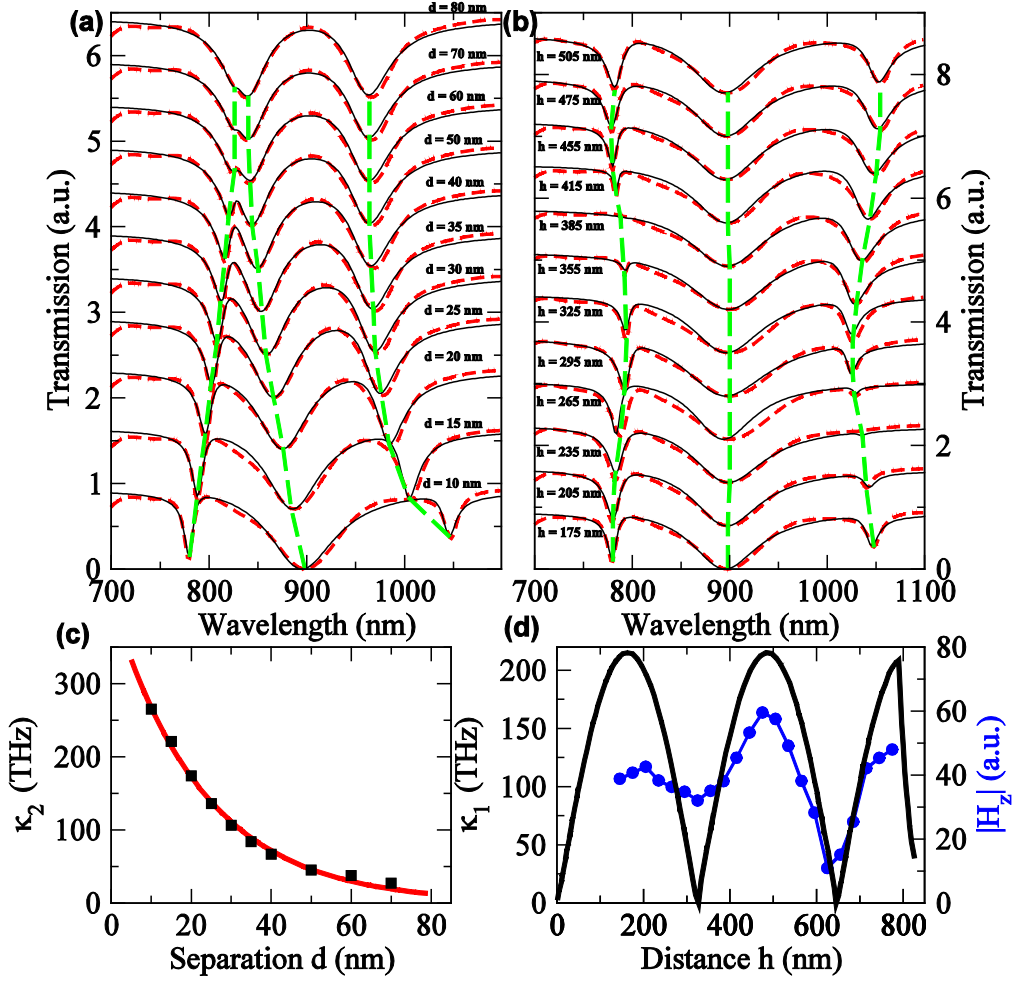

FIG. 5. Transmission by numerical simulation (dashed red) and TCMT fitting (solid black). (a) Transmission spectra as the dimer gap $d$ increases from 10 nm to 80 nm. (b) Transmission spectra as the waveguide-dimer distance $h$ increases form 175 nm to 445 nm. The individual curves are shifted vertically for clarity. Dashed green are guides for eye in both (a) and (b). (c) Fitted coupling constant $\kappa_2$ (square) as a function of the distance $d$, when $s = 10$ nm, the solid line is an exponential fit. (d) Fitted coupling constant $\kappa_1$ (circle) as a function of $h$, for $d = 10$ nm, overlapped is field intensity $|H_z|$ at $\lambda = 900$ nm ($\omega = \omega_0$) along the stub boundary in the tooth-shaped structure [see Fig. 4(h)].

by $s$ but also by $h$ because the field magnitude forms Fabry-Perot-like nodes in the stub.[13] As seen in Fig. 5(d), the fitted coupling coefficient $\kappa_1$ oscillates as a function of $h$. Interestingly,

$\kappa_1$ is approximately consistent with the periodical variation of $|H_z|$ at the resonant frequency ($\lambda = 900$ nm) in the corresponding tooth-shaped structure. More specifically, near the zero nodes of the magnetic field at $h = 327.2$ nm and 645.1 nm [see Fig. 4(h)], the coupling strength $\kappa_1$ minimizes. The side band dips can become very small to be even unrecognizable [see Fig.5(b)], for example, at $h = 235$ nm, 385 nm, and 625 nm. We stress that the dips are always observed (visible) in the lossless TCMT prediction [see Fig. 2(b)]. The reason lies in threefold: (i) the realistic system is with Ohmic loss which may make small dips even invisible; (ii) the coupling strength in the dimer would vary as $h$ changes; (iii) more strictly, the coupling strength shall be replaced by complex-valued coupling coefficient, and the resonance frequency shifts as the phases of the coefficient that also depends on wave reflection at the interfaces of the resonant cavities.[19] Finally, the oscillation periods of $\kappa_1$ and $|H_z|$ in Fig. 5(d) do not follow each other at small and large $h$, e.g., for $h < 150$ nm and $h > 750$ nm. Small $h$ would introduce direct near-field coupling between the dimer cavities and the slot waveguide. While for large $h$, the extremity and finite-width effects come into play. The amplitude of the magnetic field decreases rapidly when $h > t = 782$ nm, which does not allow sufficient power tunneling into and out of the dimer cavities.

**Conclusions**

In summary, we have investigated a phenomenon that is similar to anti-EIT in a nanometric plasmonic waveguide system with slot-stub-dimer cascaded geometry. Numerical results by FEM agree well with analytical predictions by a dynamic theory. Reminiscent to atomic physics, this coherently resonant system exhibits collective dark-states which lead to narrow spectral responses by reversing a PIT peak. As EIT can be regarded as a special case of Fano resonance, we expect more complicated dynamics and spectrum evolution if the dimer is heterogeneous, filled with nonlinear material,[20] or cascaded to additional cavities. The results could be used in designing densely integrated optical circuits, in optical sensing and switching.

**Acknowledgement** This work was supported by NSFC (11004043 and 11274083), and SZMSTP (Nos. JCYJ20120613143649014, KQCX20120801093710373, and 2011PTZZ048). J.J.X is also supported by Natural Scientific Research Innovation Foundation in Harbin Institute of Technology (No. HIT.NSRIF.2010131). The authors acknowledge assistance from the Key Lab of Terminals of IoT and the National Supercomputing Center in Shenzhen.

**References**
1. B. Luk'yanchuk, N. I. Zheludv, S. A. Maier, N. J. Halas, P. Nordlander, H. Giessen, and C. T. Chong, Nat. Mater. **9**, 707 (2010).
2. S. E. Harris, Phys. Today **50**, 36 (1997).
3. M. Fleischhauer, A. Imamoglu, and J. Marangos, Rev. Mod. Phys. **77**, 633 (2005).
4. P. Tassin, L. Zhang, Th Koschny, E. Economou, and C. Soukoulis, Phys. Rev. Lett. **102**, 187401 (2009).
5. Q. Zhang, J. J. Xiao, X. M. Zhang, Y. Yao, and H. Liu, Opt. Express **21**, 6601 (2013).
6. Q. Zhang and J. J. Xiao, Opt. Lett. **38**, 4240 (2013).
7. N. Liu, L. Langguth, T. Weiss, J. Kästel, M. Fleischhauer, T. Pfau, and H. Giessen, Nat. Mater. **8**, 758 (2009).
8. R. Taubert, M. Hentschel, and H. Giessen, J. Opt. Soc. Am. B **30**, 3123 (2013).
9. R. Taubert, M. Hentschel, J. Kastel, and H. Giessen, Nano Lett. **12**, 1367 (2012).
10. Z. Zhang, L. Zhang, H. Q. Li, and H. Chen, Appl. Phys. Lett. **104**, 231114 (2014).
11. H. Lu, X. Liu, D. Mao, Y. Gong, and G. Wang, Opt. Lett. **36**, 3233 (2011).
12. Z. R. Zhang, L. W. Zhang, P. F. Yin, and X. Y. Han, Physica B **446**, 55 (2014).
13. X. S. Lin and X. G. Huang, Opt. Lett. **33**, 2874 (2008).
14. X. Zhou and L. Zhou, Appl. Opt. **52**, 480 (2013).
15. P. K. Jha, M. Mrejen, J. Kim, C. Wu, X. Yin, Y. Wang, and X. Zhang, Appl. Phys. Lett. **105**, 111109 (2014).
16. H. A. Haus, *Waves and Fields in Optoelectronics* (Prentice Hall, Englewood Cliffs, 1984).
17. W. Suh, Z. Wang, and S. Fan, IEEE J. Quant. Electron. **40**, 1511 (2004).
18. C. W. Hsu, B. G. DeLacy, S. G. Johnson, J. D. Joannopoulos, and M. Soljačic, Nano Lett. **14**, 783 (2014).
19. J. S. White, G. Veronis, Z. Yu, E. S. Barnard, A. Chandran, S. Fan, and M. L. Brongersma, Opt. Lett. **34**, 686 (2009).
20. Y. Xu, X. Wang, H. Deng, K. Guo, Opt. Lett. **39**, 5846 (2014).